\documentclass[]{spie}

\usepackage[pdftex]{graphicx}
\usepackage[usenames,dvipsnames]{xcolor}
\usepackage{fancyhdr}
\usepackage{amssymb}
\usepackage{amsmath}
\usepackage{hyperref}
\usepackage{url}
\usepackage{rotating}
\definecolor{DarkRed}{RGB}{128,0,0}
\definecolor{DarkGreen}{RGB}{0,16,0}
\definecolor{DarkBlue}{RGB}{0,0,16}
\definecolor{DarkCyan}{RGB}{0,128,128}
\definecolor{DarkMagenta}{RGB}{32,0,32}
\definecolor{Grey}{RGB}{32,32,32}
\hypersetup{
    unicode=false,             
    pdftoolbar=false,          
    pdfmenubar=true,           
    pdffitwindow=false,        
    pdfstartview={FitH},       
    pdftitle={Techniques for Radio-Isotope Identification},  
    pdfauthor={Hague, E. J.},  
    pdfsubject={Radiation Detection},  
    pdfcreator={Hague, E. J.}, 
    pdfproducer={DOE; RSLA},  
    pdfkeywords={Radiation Detection} {Isotope Identification} {Data Science} {Machine Learning} {Neural Network} {Template Matching} {Likelihood} {Spectroscopy},  
    pdfnewwindow=true,         
    colorlinks=true,           
    linkcolor=DarkBlue,        
    citecolor=DarkGreen,       
    filecolor=DarkMagenta,     
    urlcolor=DarkCyan          
}

\title{A Comparison of Adaptive and Template Matching Techniques for Radio-Isotope Identification} 
\author{
  Emma J. Hague\supit{a}, 
  Mark Kamuda\supit{b}, 
  William P. Ford\supit{a}, 
  Eric T. Moore\supit{a}, and 
  Johanna Turk\supit{c}
\skiplinehalf
\supit{a}Remote Sensing Laboratory, Joint Base Andrews, Maryland\\
\supit{b}University of Illinois, Champaign-Urbana, Illinios\\
\supit{c}Barnstorm Research, Boston, MA}

\authorinfo{
Send correspondence to Emma Hague: E-mail: hagueej@nv.doe.gov, Telephone: +1.301.817.3361
}

\begin{document}
\maketitle 

\begin{abstract}
  We compare and contrast the effectiveness of a set of adaptive and non-adaptive algorithms
  for isotope identification based on gamma-ray spectra. One dimensional energy spectra are
  simulated for a variety of dwell-times and source to detector distances in order to reflect
  conditions typically encountered in radiological emergency response and environmental monitoring
  applications. 
  We find that adaptive methods are more accurate and computationally efficient than non-adaptive
  in cases of operational interest.
\end{abstract}


\section{INTRODUCTION}
\label{sec:intro}
Algorithmic gamma-ray spectrum classification and isotope identification is a well developed
field\cite{kamudajifu2018,Sharma12,Ford18,Moore19,1DSDRD,CNNSDRD,sorma18}. 
In radiological search and environmental health operations there are widely deployed tools based on
non-adaptive template matching schemes to or supplement a human spectroscopist or even fully 
automate spectrum classification and isotope identification. 
In recent years adaptive techniques, such as neural networks and decision tress, have been
developed in the research literature\cite{kamudajifu2018,Sharma12}. 
These new techniques have not been directly compared to the historical approaches in order to gauge
any increased accuracy or efficiency. 
Furthermore, there is a dearth of study pertaining to the often crucially important operational
considerations of dwell-time for collecting each spectrum and the distance between the source and
the detector. 
Both of these operational factors can dramatically influence the classification accuracy and, thus,
yield significantly different operational outcomes. 
In this work, we endeavor to systematically compare two non-adaptive (template matching) 
techniques to more modern adaptive techniques in the context of realistic operational time and
distance, and comment on the befits of each. 

\section{DATA}
\label{sec:data}
All data in this work are derived from modeled source and background
templates, and have been Poisson sampled to generate sufficient training and testing sets. 
An advantage of using modeled data in this study is that we can precisely control, and know in
advance, the {\em true} signal strength of each modeled scenario. This knowledge lends confidence
and credibility to our conclusions since we will not be guessing, or relying on human
spectroscopists, for the truth of each encounter. 
Another advantage of using modeled data is that we can avoid the class imbalance 
problems\cite{Sharma12,} typically
associated with data sets collected from controlled laboratory or operational settings. 
For Example, even though
$^{67}$Cu is a much less common isotope to encounter in the environment, we will have just as many
examples of it as we will the ubiquitous $^{99m}$Tc.
An major disadvantage is that modeling software is known to not fully and correctly reproduce the
myriad of complications that are routinely found in real-world data. 
Towards our goal of directly comparing and contrasting adaptive and non-adaptive methods, 
we asses that the advantages out-weigh the draw backs.

\subsection{Templates} 
\label{sec:template-data}
Spectral templates are crated using the Gamma Detector Response and Analysis Software
(GADRAS)\cite{Gadras}.  The modeled detector is a 2"$\times$4"$\times$16" NaI(Tl) crystal with 1024
channels in energy.  GADRAS generates these templates using historical in-situ data collected in
a laboratory setting or, as in the case of geographically specific backgrounds, in the field.
The collected data are then convolved with the known detector response functions to produce an
asymptotic spectral shape.  This shape is asymptotic in the sense that it is modeled as a very
long dwell measurement -- in our case 24 hours -- which is nearly free of channel-wise Poisson
noise\footnote{
Measurement noise is still evident at the highest energies, $\gtrsim 2.75$MeV, but this range does
not contain any relevant spectral information.
}. 
These shapes are then scaled to the dwell-time being studied and employed both for
ensemble generation and in the non-adaptive template matching techniques described below.

The templates are generated with 1024 channels, spanning the gamma-ray energy range from zero to
3 MeV.  The width in energy of each channel is $\approx2.93$ keV. 
The first ten bins are ignored for all of the analysis in order to approximate low-level 
energy discriminators often found in fielded detector systems. 
Thus, the gamma-ray energy spectra are divided into 1014 energy bins from 29.297 keV to 3 MeV.

\begin{figure}[ht]
  \begin{center}
    \includegraphics[width=0.80\linewidth]{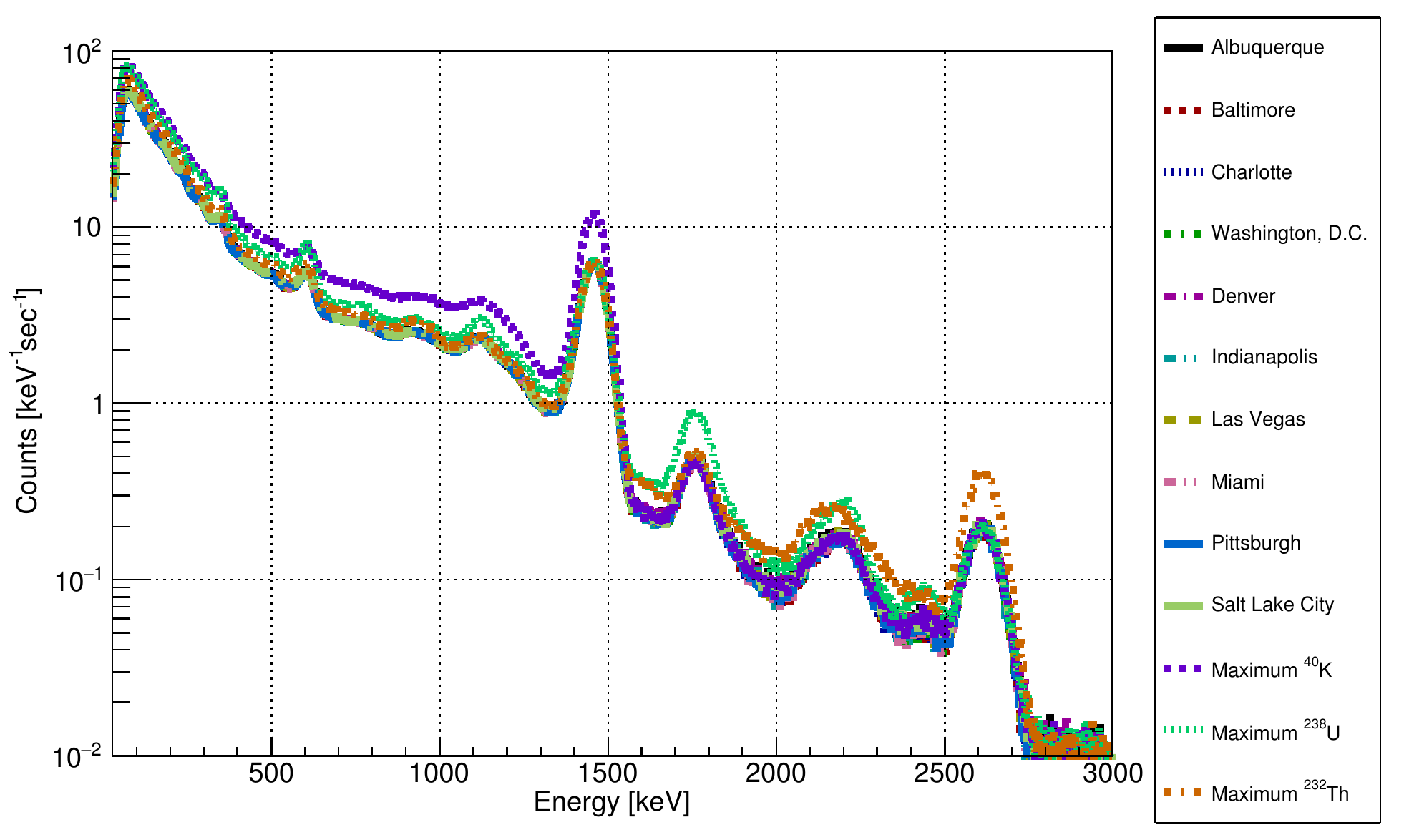}
  \end{center}
  \caption[example]{
    \label{fig:bkg-templs}
    Modeled background spectral templates.  The gamma-ray spectrum typical of 10 cities in the
    United States.  Additionally, we generate background templates where the ratios of each of
    the naturally occurring radio-isotopes are maximized with respect to it's natural range.
  }
\end{figure} 

In Figure \ref{fig:bkg-templs}, we show the background templates typical of ten different cities 
in the United States. 
Additionally, we generate background templates where the ratios of each of the naturally occurring
radio-isotopes $^{40}$K, $^{238}$U, and $^{232}$Th are maximized with respect to the range over
which they are commonly found in the natural environment. 
This variety in background templates is included in this study in an attempt to model some
of the variability commonly found in operational or in situ measurements, and thus potentially
increase the robustness of the methods developed and studied. 
We note, however, that the difference in spectral shape between the different background templates
are insufficient to fully reflect operational concerns. For example, even the
$^{40}$K$_{\text{}max}$ background is not as deviant as what might be found after a heavy rain
leaches $^{222}$Rn from the soil\cite{Sharma12}.

\begin{figure}[ht]
  \begin{center}
    \begin{tabular}{ccc}
      \includegraphics[width=0.80\linewidth]{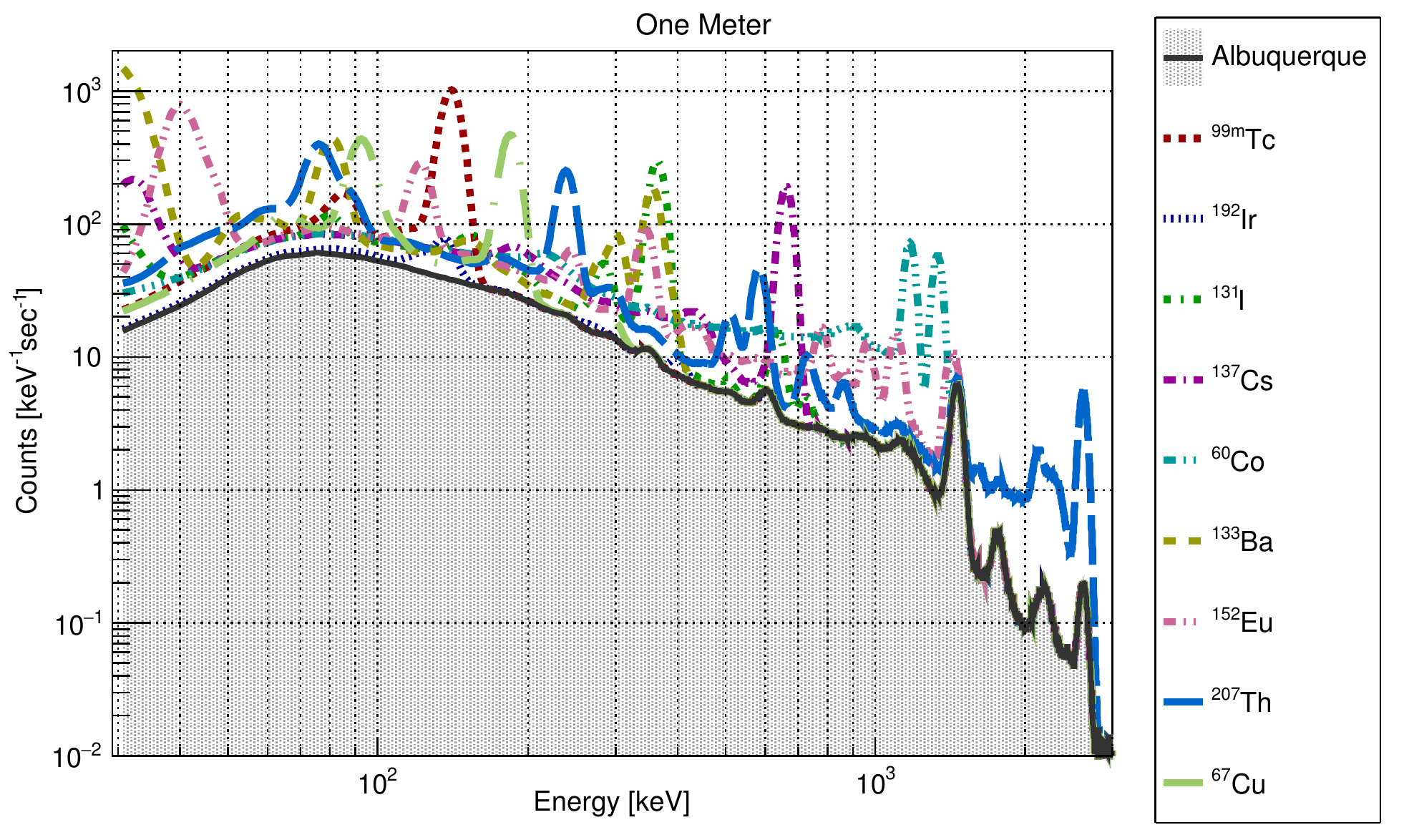} \\
      \includegraphics[width=0.80\linewidth]{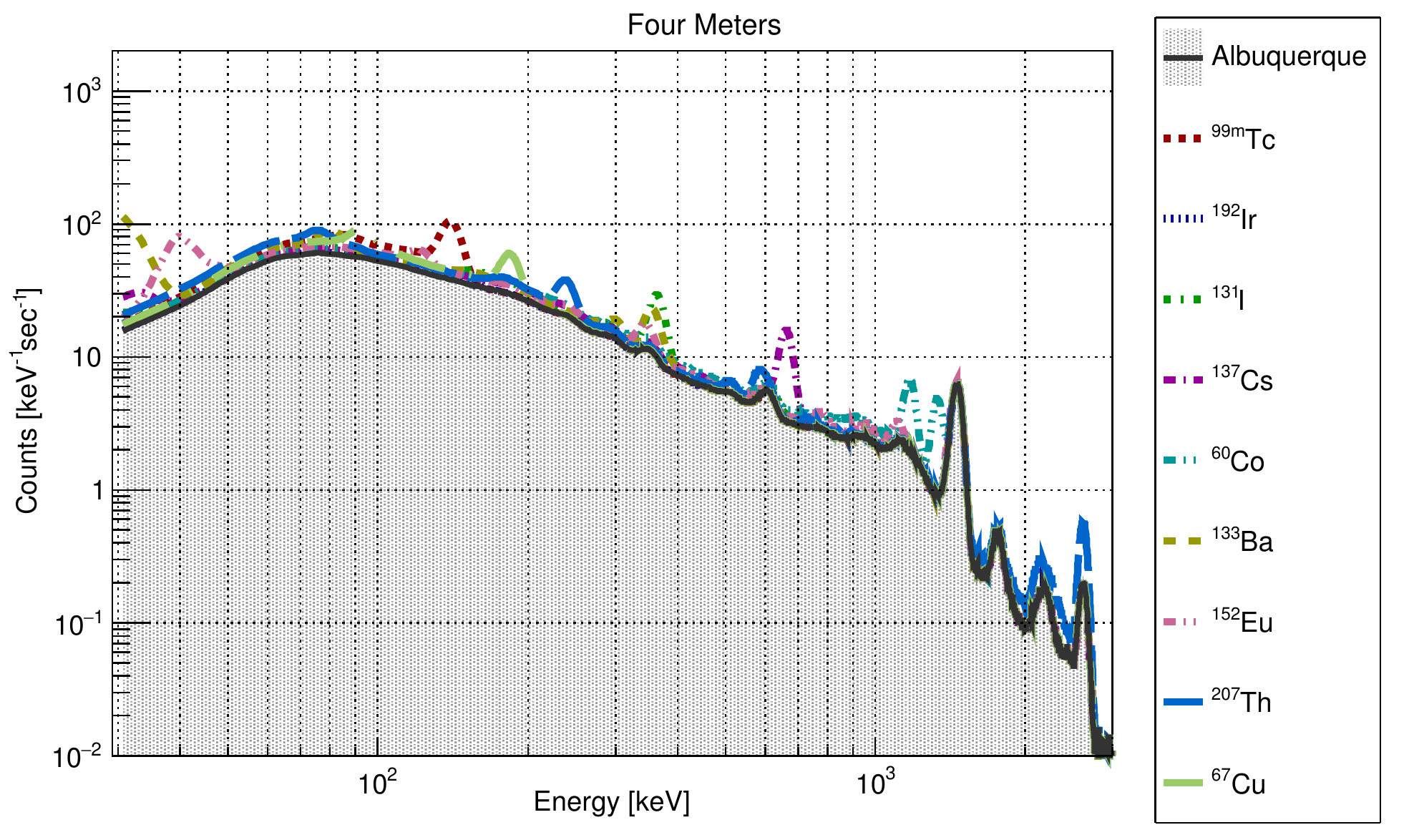}
    \end{tabular}    
  \end{center}
  \caption[example]{
    \label{fig:src-tmpls} 
    Modeled source spectral templates, when combined with the background typical of Albuquerque,
    NM. Note that both the vertical axis (Counts [keV$^{-1}$sec$^{-1}$]) and horizontal axis
    (Energy [keV$^{-1}$]) are logarithmically scaled in order to highlight the low-energy
    channels where most of the isotopic spectral peaks are present.  The top panel shows the
    modeled detector response with one meter of separation between each source and the detector,
    and the bottom shows that with 4 meters of separation.  
    The visible difference between the top and bottom panels illustrates the significant effect 
    of atmospheric attenuation will have on observed gamma-ray energy spectra.
  }
\end{figure} 

Nine source (i.e. signal) templates are also modeled and are listed in Table
\ref{table:isotopes}. The modeled isotopes were chosen to be representative of those commonly used
in medical and industrial applications.  The modeled activity of each isotope is roughly that often
found in medicine and industry, but was further adjusted to give a (mostly) clear signal at short
distances and medium dwell-times, as shown in the third column of Table \ref{table:isotopes}. 

The detector response was modeled with the distance between the source and detector 
$d \in \{1, 2, 4, 5, 6, 8\}$ meters, and dwell-times 
$t \in \{0.25, 0.5, 1, 2, 4, 8, 16, 32, 64, 128, 256\}$ seconds. 
This range of distances and dwell-times is typical of those found in radiation health and safety 
applications. 
The source-to-detector distance induces significant decrease in signal strength due to gamma-ray 
attenuation in air. 
The last two columns of Table \ref{table:isotopes} show the effect of attenuation on the signal 
strength. This attenuation, and the general shape of the source templates when combined with a 
typical background, Albuquerque, is plotted in Figure \ref{fig:src-tmpls} for the distances 
one meter and four meters.

\begin{table}[ht]
  \begin{center}
    \begin{tabular}{|c|c|c|c|c|}
      \hline
      Isotope & $N_{s}$ [sec$^{-1}$] & $N_{s}/N_{b}$ [1 m] & $N_{s}/N_{b}$ [4 m] & $N_{s}/N_{b}$ [8 m] \\
      \hline
      $^{99m}$Tc & 6756.72 & 1.45 & 0.18 & 0.05 \\
      $^{192}$Ir & 422.314 & 0.09 & 0.01 & 0.003 \\
      $^{131}$I  & 5702.31 & 1.23 & 0.15 & 0.04 \\
      $^{137}$Cs & 6478.99 & 1.39 & 0.15 & 0.04 \\
      $^{60}$Co  & 8636.7  & 1.86 & 0.21 & 0.05 \\
      $^{133}$Ba & 8543.61 & 1.84 & 0.20 & 0.05 \\
      $^{152}$Eu & 8151.16 & 1.75 & 0.19 & 0.05 \\
      $^{207}$Th & 7596.29 & 1.63 & 0.20 & 0.05 \\
      $^{67}$Cu  & 5657.2  & 1.22 & 0.16 & 0.04 \\
      \hline
    \end{tabular}
    \caption{
      \label{table:isotopes}
      The simulated isotopes considered in this study. 
      In the second column we show the number of signal events, $N_{s}$, for a dwell-time of one
      second and at a distance of one meter. 
      In the later columns we show the ratio of expected signal counts to expected background counts
      $N_{s}/N_{b}$ at one, four, and eight meters. 
      This table illustrates the total number of counts associated with each modeled isotope and
      the diminishing signal strength at increasing distances.   
    }
    
  \end{center}
\end{table}

\subsection{Ensemble Generation} 
\label{sec:ensemble-data}
Ensembles of simulated data were generated for the purposes of both training and testing the
various algorithms studied here. 
For each combination of 13 backgrounds, 9 isotopes, 11 dwell-time and 6 distances a simulated
spectrum is generated by Poisson sampling the appropriate template. 
This sampling is repeated $1\times10^{3}$ times for each of two ensembles which we label
``training'' and ``testing'' receptively. 
Thus each ensemble contains $7.722\times10^{6}$ total simulated spectra. 
Within the two ensembles, each distance, dwell-time, and isotope combination contains 
$1.3\times10^{4}$ spectral manifestations.

\section{METHODS}
\label{sec:methods}

\subsection{Template Matching} 
\label{sec:template-matching}
Template matching techniques are ubiquitous in gamma-ray spectroscopy and have historically been
used since the underlying mathematics required to implement them has been available for many
decades\cite{cernroot,roofit,maxlike}. 
These techniques are non-adaptive in the sense that they are designed to choose the
model that best represents the data from a collection of predefined choices, without prior
access to any data; they need not be trained. 

We can express the PDF of the observed foreground spectrum as a function of gamma-ray energy, $E$,
as a linear combination of the background template and the isotope (i.e. signal) template,
\begin{equation}
  f(E|n_{b}, n_{s}) = n_{b}f_{b}(E) + n_{s}f_{s}(E), \label{equ:pdf}
\end{equation}
where $n_{b}$ and $n_{s}$ are the fit parameters associated with the number of background and
signal events, respectively. The functions $f_{b}(E)$ and $f_{s}(E)$ represent the background and
signal templates, respectively, normalized such that they each integrate to unity over the range 
29.297 keV to 3 MeV.

A binned likelihood objective function is constructed, using the standard approach\cite{maxlike}, 
from the Poisson probability of observing $N_{i}$ counts in a bin, 
\begin{equation}
  \mathcal{L}(n_{b}, n_{s}) = \prod_{i=1}^{1014}
  \left(f(E_{i}|n_{b}, n_{s})\right)^{N_{i}}e^{-f(E_{i}|n_{b}, n_{s})}/N_{i}!, \label{equ:like}
\end{equation}
where $E_{i}$ is the energy of each bin and $N_{i}$ is the observed simulated counts in each bin.
For numerical effectiveness the monotonic property of the logarithm can be exploited, and
non-parameter dependent terms dropped, such that the actual function to be minimized with respect 
to the parameters $n_{b}$ and $n_{s}$ is 
\begin{equation}
  \Lambda(n_{b}, n_{s}) \equiv -\ln \mathcal{L}(n_{b}, n_{s})
  \equiv \sum_{i=1}^{1014} f(E_{i}|n_{b}, n_{s}) - N_{i} \ln f(E_{i}|n_{b}, n_{s}). \label{equ:log-like}
\end{equation}

If, and only if, the number of observed counts in each bin is large enough one may minimize 
the $\chi^{2}$ objective function,
\begin{equation}
  \chi^{2}(n_{b}, n_{s}) = 
  \sum_{i=1}^{1014} \frac{\left(N_{i} - f(E_{i}|n_{b}, n_{s})\right)^{2}}{f(E_{i}|n_{b}, n_{s})}. 
  \label{equ:chi-sqr}
\end{equation}
For most of the cases under consideration in this work, and indeed most of the cases encountered in
radiation and environmental safety, the number of counts in {\em all} bins is insufficient to
justify the use of this objective function. We include it's use in this work, however, since this
objective function is commonly used in the fielded applications and as another metric against which 
we can compare the newer, adaptive, techniques.

Once the objective function(s) is minimized with respect to the free parameters, we choose the
isotope template that best represents the data by calculating the Kolmogorov-Smirnov probability 
for each with respect to the simulated binned data under consideration. We further restrict the
choice of best fit template by not allowing the fraction of fit signal events, $n_{s}$, to be 
less than 5\%, since any signal that is sufficiently close to zero would be indistinguishable from
background.

\subsection{Adaptive Techniques} 
\label{sec:adaptive}

The two adaptive methods explored in this work are dense neural networks (DNN) and convolution
neural networks (CNN). 

\begin{figure}[h]
  \begin{center}
    \begin{tabular}{cc}
      \includegraphics[width=0.48\linewidth]{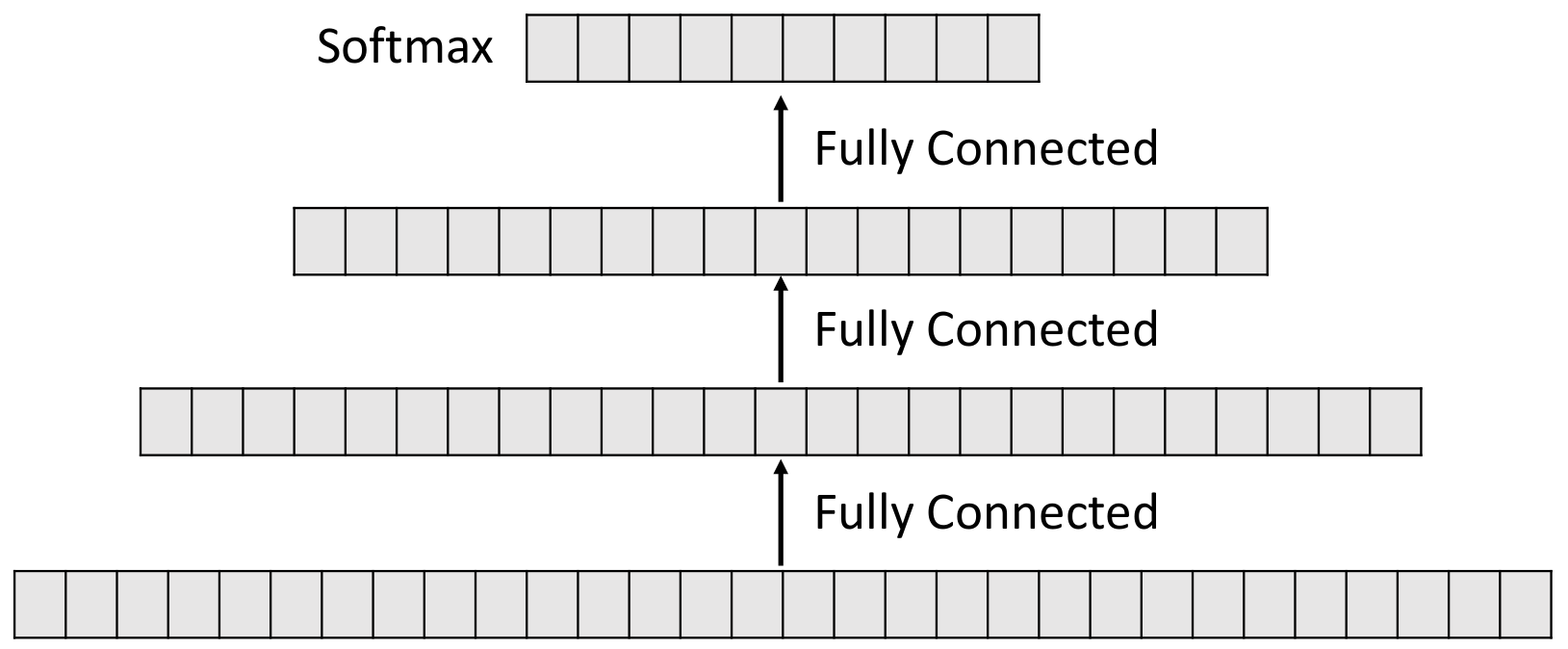} & 
      \includegraphics[width=0.48\linewidth]{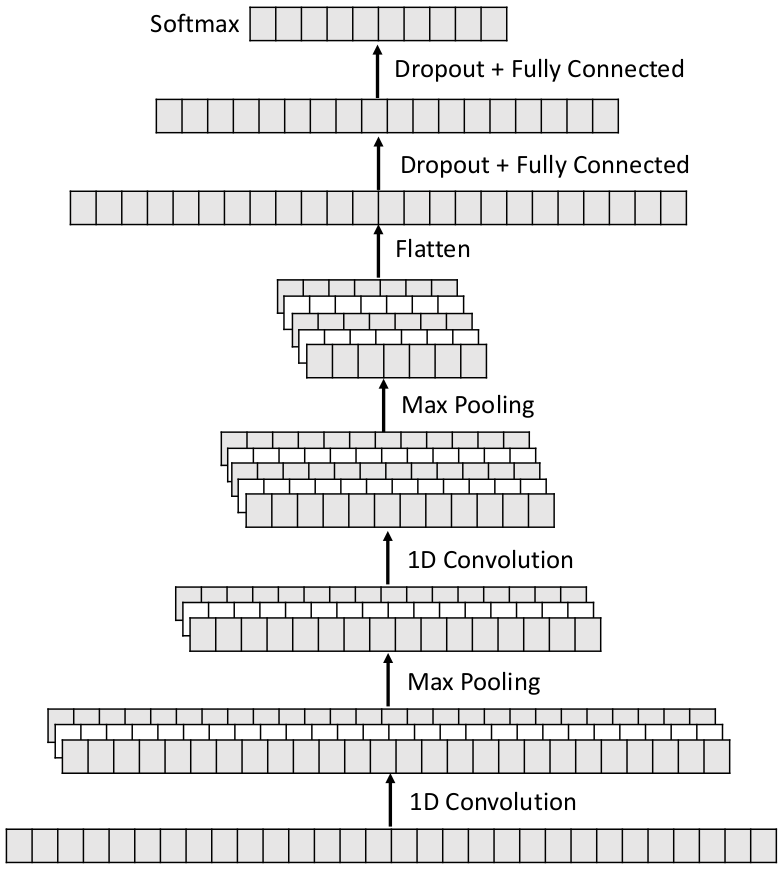}
    \end{tabular}    
  \end{center}
  \caption[example]{
    \label{fig:nn-examples}
    Left: An example representation of a dense neural network (DNN) where the input layers are the
    counts in each energy bin of the spectrum, the intervening layers are fully connected,
    and the output layer is softmax for determining the most likely isotope.
    In the right panel we show a deeper, more complicated convolutional neural network (CNN) where
    the intervening layers are successive combinations of pooling, one dimensional convolutions,
    flattening, and finally dropout and fully connected layers. 
    These representations are drawn from Kamuda, et al.\cite{kamudajifu2018}
  }
\end{figure} 

An example DNN architecture with two hidden layers is shown in 
the left panel of Figure \ref{fig:nn-examples}. 
In this network, each channel of a gamma-ray spectrum is connected to each
node in the next layer. These dense connections continue for each layer in the network.

An example of a CNN is shown in Figure \ref{fig:nn-examples}. The input and output of the CNN are
the same as the DNN. The difference between the CNN and DNN are the convolution and max pooling
layers. Convolution layers activations are created by convolving 1-D filters with the previous
layer's signal. Max pooling is a sub-sampling operation that attempt to combine low-level features
and to encourage spatial invariance by reducing the resolution of the previous layers
\cite{Scherer2010}. After the convolution and pooling layers, the features are flattened into a
vector and fed into a fully-connected architecture. The weights of the fully-connected network and
the 1-D convolution filters are learned through training.

In general, neural networks have a tendency to memorize their training set in a process called
overtraining. An overtrained neural network will tend to incorrectly identify novel data. To
prevent this, regularizing hyperparameters were used in the to prevent overfitting and optimize
performance. There is currently no known method to know which hyperparameters have an impact on
model performance before training. Because of this, a number of hyperparameters are typically added
to a model and a random hyperparameter search is used to identify those which are important
\cite{Bergstra2012}. 

Ranges of hyperparameters explored for the DNN and CNN are shown in 
the left and right panels of Table \ref{table:nn-params}, respectively. 
A total of 128 networks
were trained with randomly chosen hyperparameters from these ranges. The networks were trained on
spectra simulated with a source-detector distance of 4 meters. The performance of each network was
compared on a separate validation dataset of spectra simulated with the same source-detector
distance. In each of these datasets 10 different spectra were simulated for each isotope and
integration time. The hyperparameter combination that performed best on this dataset was chosen as
the final training values. From these values new networks were trained using spectra simulated with
source-detector distance of 4 meters. 

\begin{table}[h]
  \begin{center}
    \begin{tabular}{cc}
    \begin{tabular}{|c|c|c|}
      \hline
      & \begin{tabular}[c]{@{}c@{}}Range \end{tabular} & 
        \begin{tabular}[c]{@{}c@{}}Final\\ Value \end{tabular} \\ \hline
        \begin{tabular}[c]{@{}c@{}}Number  of Layers \end{tabular} &  1 - 2 & 1 \\ \hline
        \begin{tabular}[c]{@{}c@{}}Number of Neurons\\ in Each Hidden Layer \end{tabular} & 
          16 - 1024 & 128 \\ \hline
        \begin{tabular}[c]{@{}c@{}}Initial\\ Learning Rate\end{tabular} & 
          10$^{-4}$- 10$^{-1}$ & 1.1 x $10^{-4}$ \\ \hline
        \begin{tabular}[c]{@{}c@{}}L2 Regularization\\ Strength\end{tabular} & 
          10$^{-2}$- 10$^{-0.5}$ & 0.24 \\ \hline
        Neuron Dropout Rate &  0.0 - 1.0 & 0.86 \\ \hline
        Batch Size &  16, 32, 64, 128 & 64  \\ \hline
        Activation Function & relu, tanh & tanh \\ \hline
        Input Scaling &  log1p, sqrt & sqrt  \\ 
        \hline
    \end{tabular}
    &
    \begin{tabular}{|c|c|c|}
      \hline
      Hyperparameter & \begin{tabular}[c]{@{}c@{}}Range\end{tabular} & 
        \begin{tabular}[c]{@{}c@{}}Final\\ Value\end{tabular} \\ \hline
          Number of Filters & 4, 8, 16, 32 & 32 \\ \hline
          \begin{tabular}[c]{@{}c@{}}Number of \\ Convolution Layers\end{tabular} & 
            1, 2 & 1 \\ \hline
            Filter Length & 2, 4, 8, 16, 32 & 16 \\ \hline
            \begin{tabular}[c]{@{}c@{}}Initial \\ Learning Rate\end{tabular} & 
              10$^{-4}$- 10$^{-2}$ & 1.3 x 10$^{-3}$ \\ \hline
              Batch Size & 16, 32, 64, 128 & 64 \\ \hline
              Dense Layers & 1 - 3 & 1 \\ \hline
              Nodes in each Layer & 10 - 512 & 14 \\ \hline
              \begin{tabular}[c]{@{}c@{}}Dense Layer \\ L2 Regularization\\ Strength\end{tabular} & 
                10$^{-3}$- 10$^{-0.5}$ & 0.015 \\ \hline
                \begin{tabular}[c]{@{}c@{}}Dense Layer \\ Neuron Dropout Rate\end{tabular} & 
                  0 - 1 & 0.19 \\ \hline
                  Activation Function & relu, tanh & tanh \\ \hline
                  Input Scaling & log1p, sqrt & sqrt \\
                  \hline
    \end{tabular}
    \end{tabular}    

    \caption{
      \label{table:nn-params}
      Left: Range of hyperparameter values for the DNN tested in a random search optimization.
      Right: Range of hyperparameter values for the CNN tested in a random search optimization.
      A total of 128 networks were trained with randomly chosen hyperparameters from these ranges,
      and the final values were used in for the results of this study.
    }    
  \end{center}
\end{table}

\section{RESULTS}
\label{sec:results}
There are many factors for evaluating the effectiveness of isotope classification and different
end-users will give different weights to each consideration. Out perspective is primarily an 
{\em operational} one; we are interested in not only class-wise accuracy but also more subtle
factors like computational efficiency and false-alarm rate (fall-out).

It is well understood that, once properly trained, adaptive techniques are highly computationally
efficient to deploy, since they rely on simple operations such as matrix multiplication and
functional computation and require no {\em a posteriori} minimization. We found this to be true
here: while the temple matching techniques routinely required multiple hours to complete (even when
using state of the art software\cite{roofit} and running parallel-ized on 8 CPU cores), 
the neural networks give results in a matter of seconds.

\begin{figure}[ht]
  \begin{center}
    \begin{tabular}{ccc}
      \includegraphics[width=0.49\linewidth]{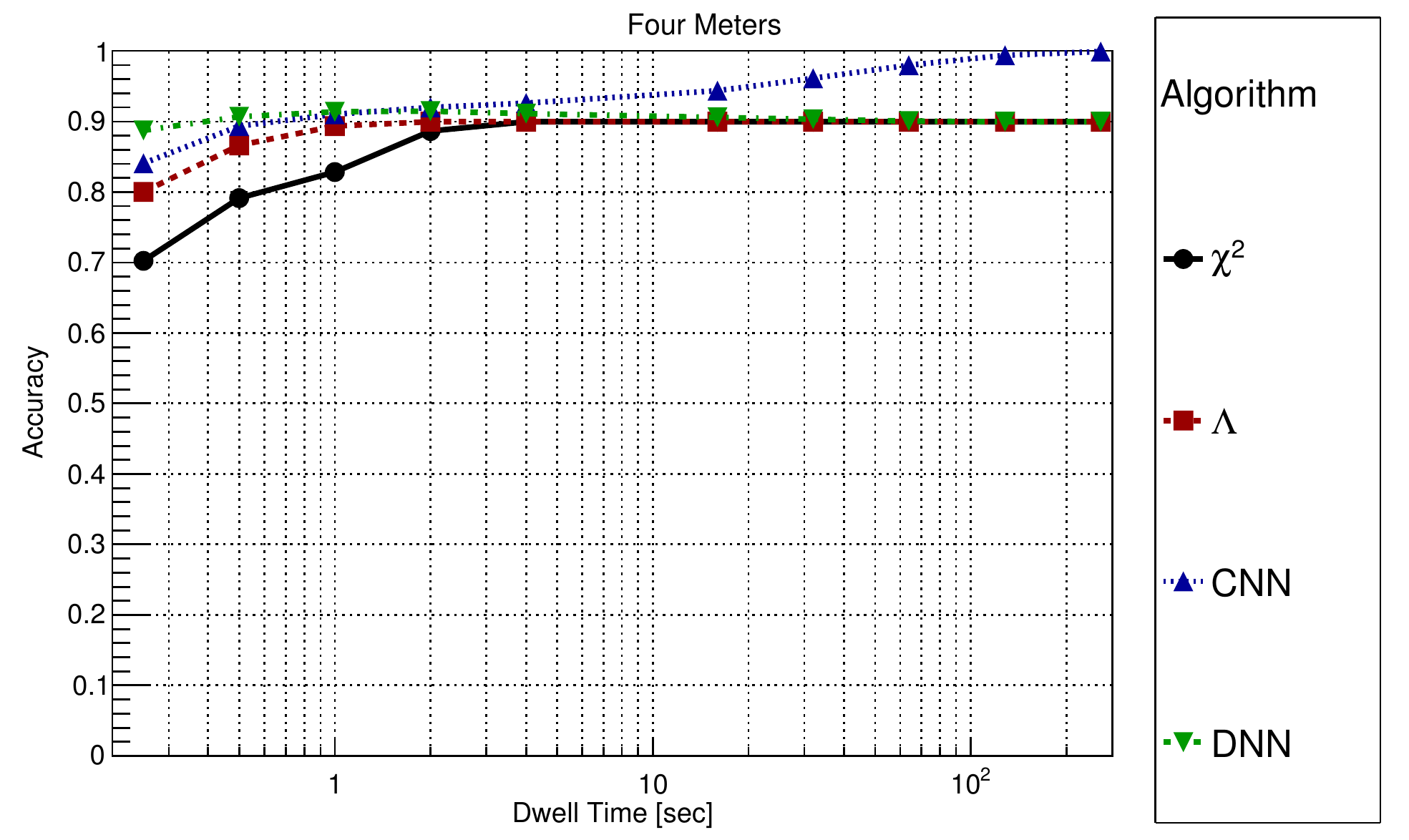} &
      \includegraphics[width=0.49\linewidth]{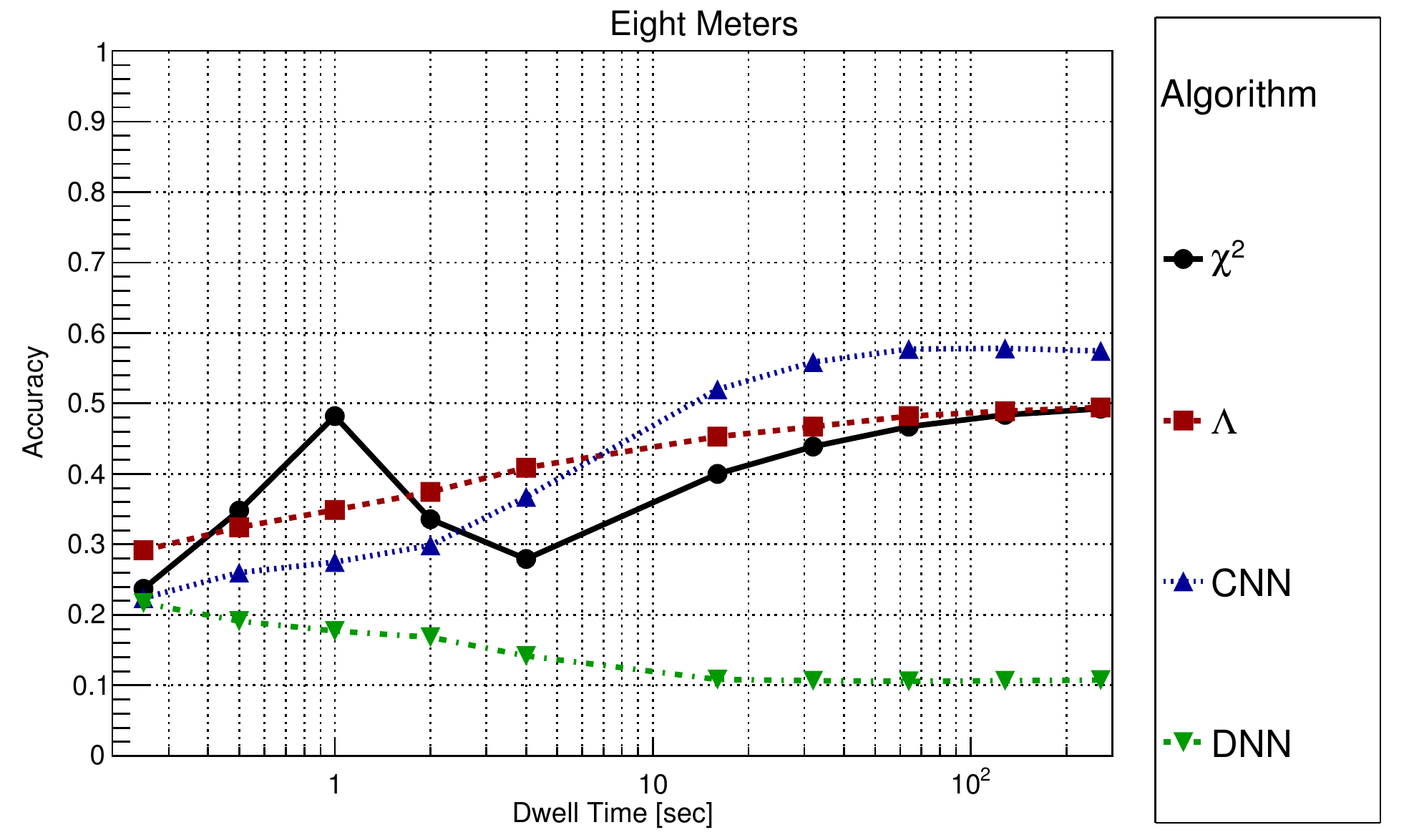} \\
    \end{tabular}    
  \end{center}
  \caption[example]{
    \label{fig:AccuracyD} 
    The accuracy of each method as a function of dwell-time at a distance of 4 meters (left) and 8
    meters (right).
  }
\end{figure}

\begin{figure}[ht]
  \begin{center}
    \begin{tabular}{ccc}
      \includegraphics[width=0.49\linewidth]{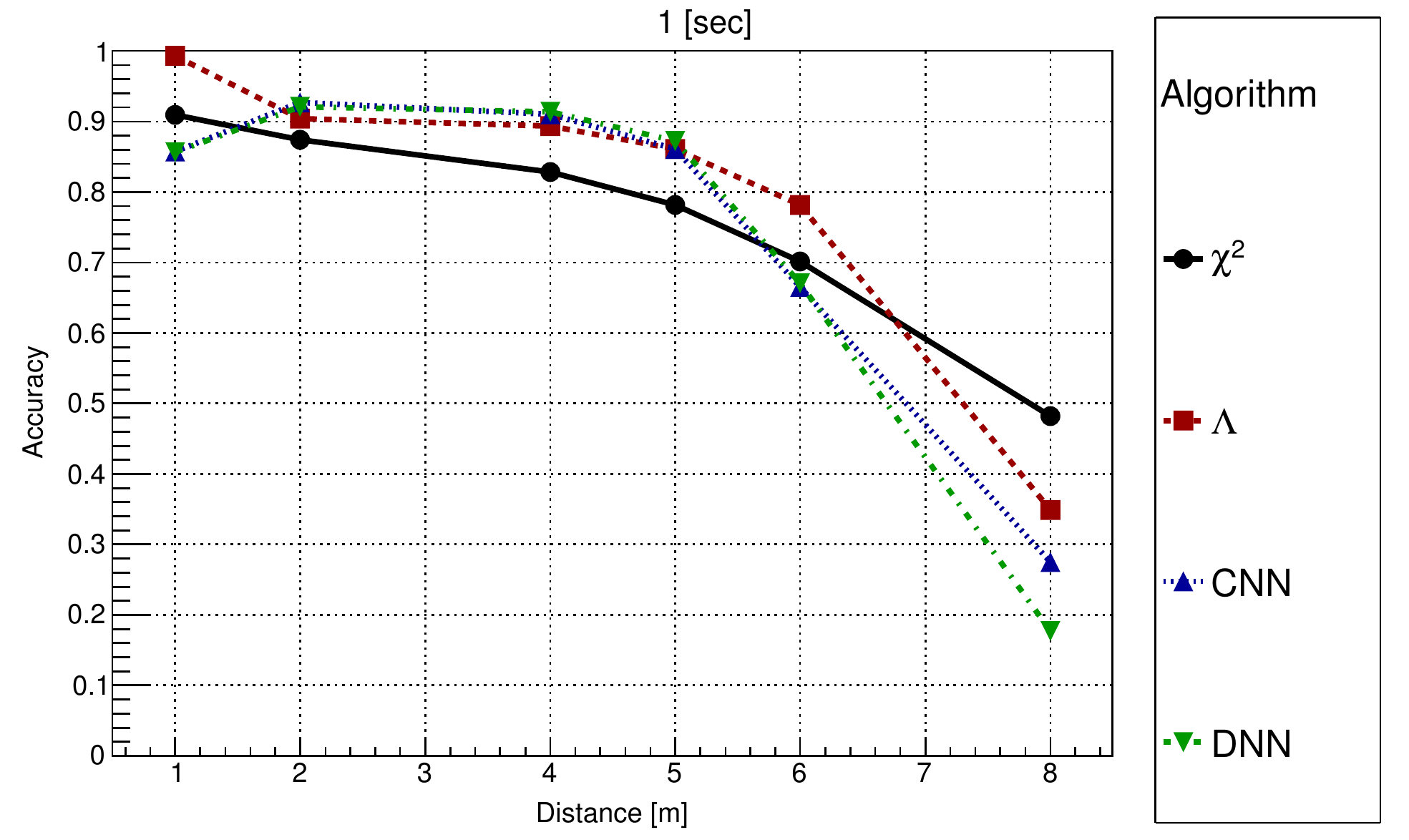} &
      \includegraphics[width=0.49\linewidth]{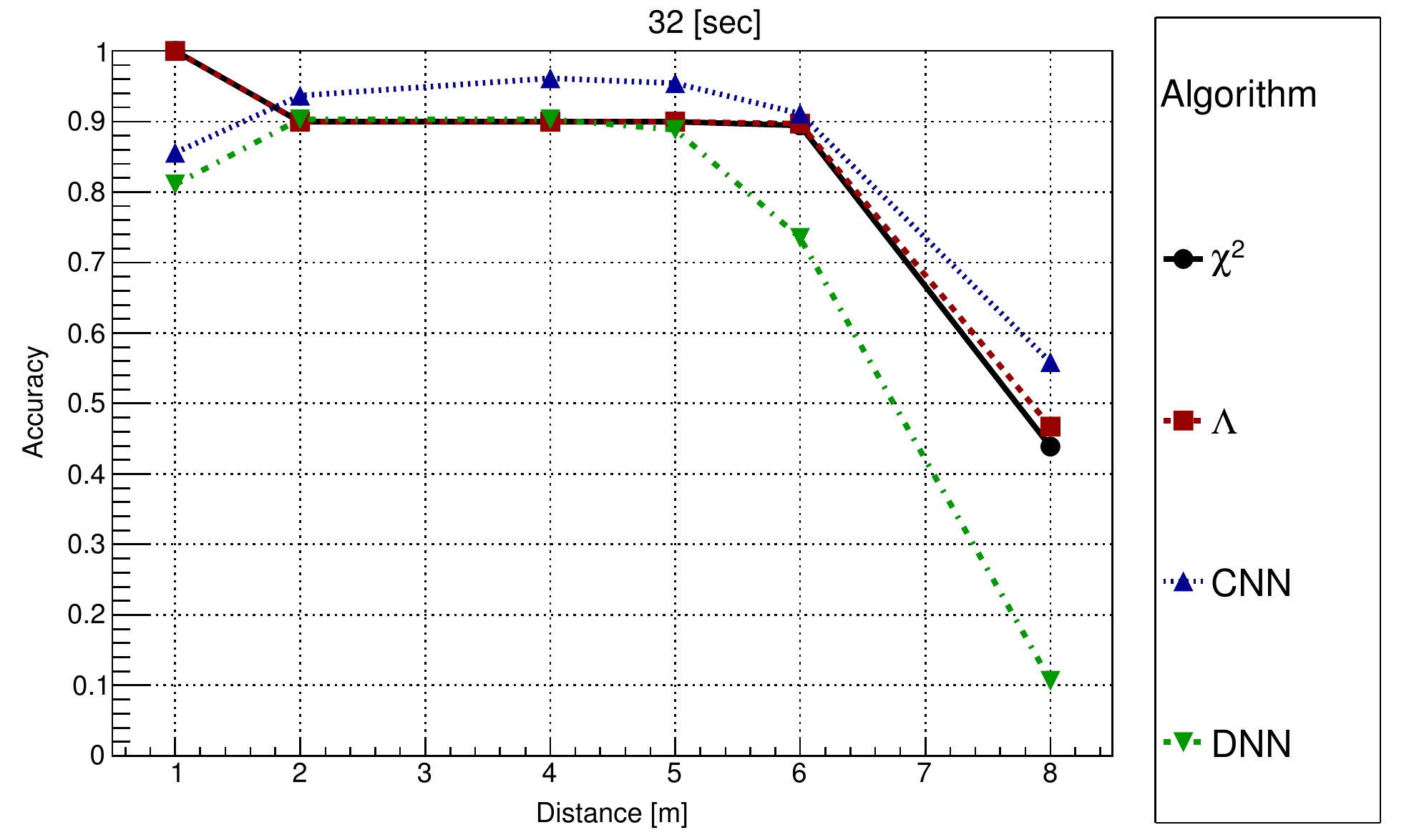} \\
    \end{tabular}    
  \end{center}
  \caption[example]{
    \label{fig:AccuracyT} 
    The accuracy of each method as a function of distance for a dwell-time 1 second (left) and 32
    seconds (right).
  }
\end{figure}

Our primary results, in terms of isotopic class accuracy, are shown in Figures \ref{fig:AccuracyD}
and \ref{fig:AccuracyT}, which illustrate the effect of increasing dwell-time and distance. 
We note that in all cases the DNN performs comparably, or better, than the other methods
considered. 
Furthermore, in the low-statistics regimes of short dwell-times and medium to long distance, the
likelihood approach performs notably better than the $\chi^{2}$  objective function. 
Based on this evidence alone it seems reasonable to suggest that currently used software packages
might consider switching to using the log-likelihood objective function.  

\begin{sidewaysfigure}[ht]
  \begin{center}
    \begin{tabular}{ccc}
      \includegraphics[width=0.48\linewidth]{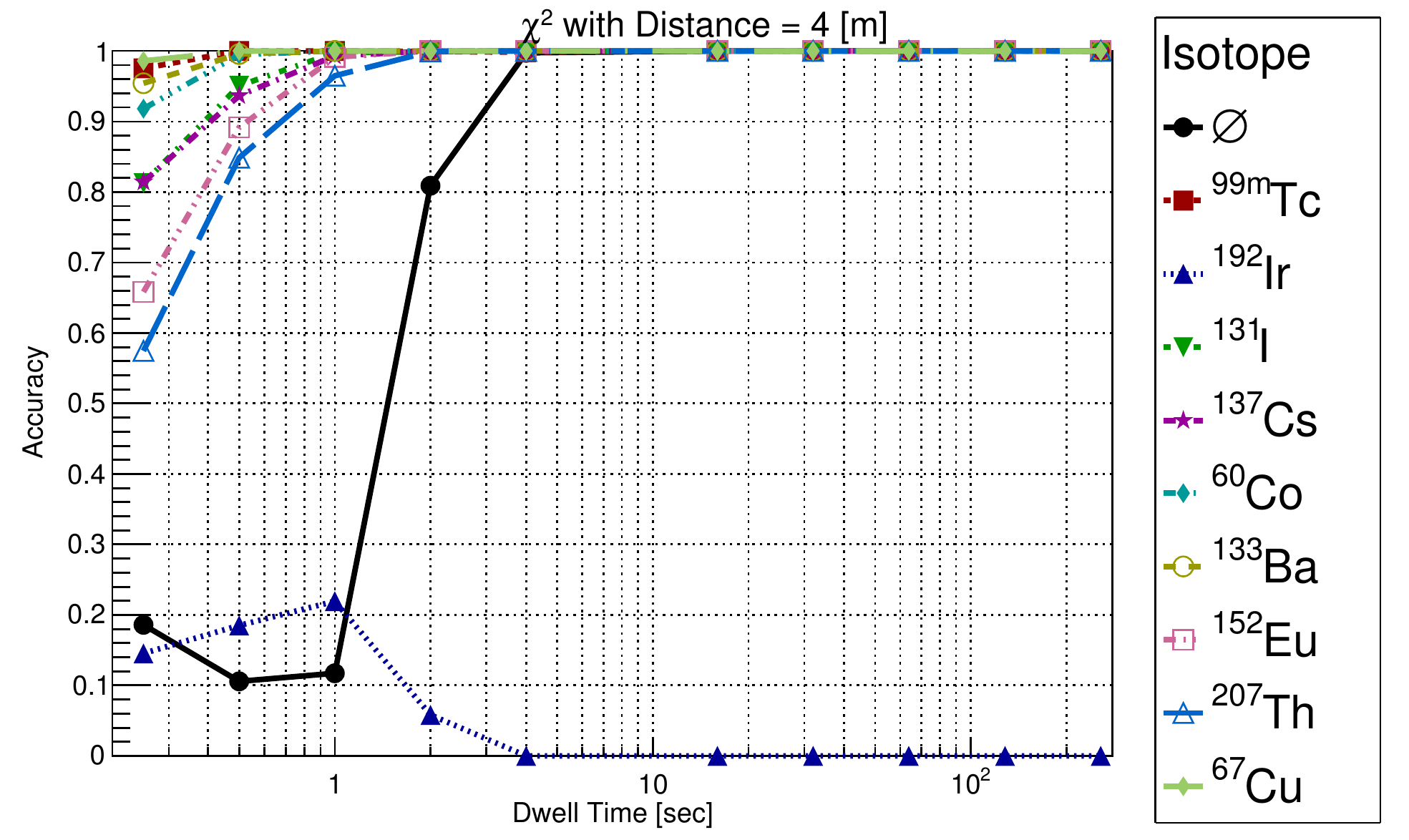} &
      \includegraphics[width=0.48\linewidth]{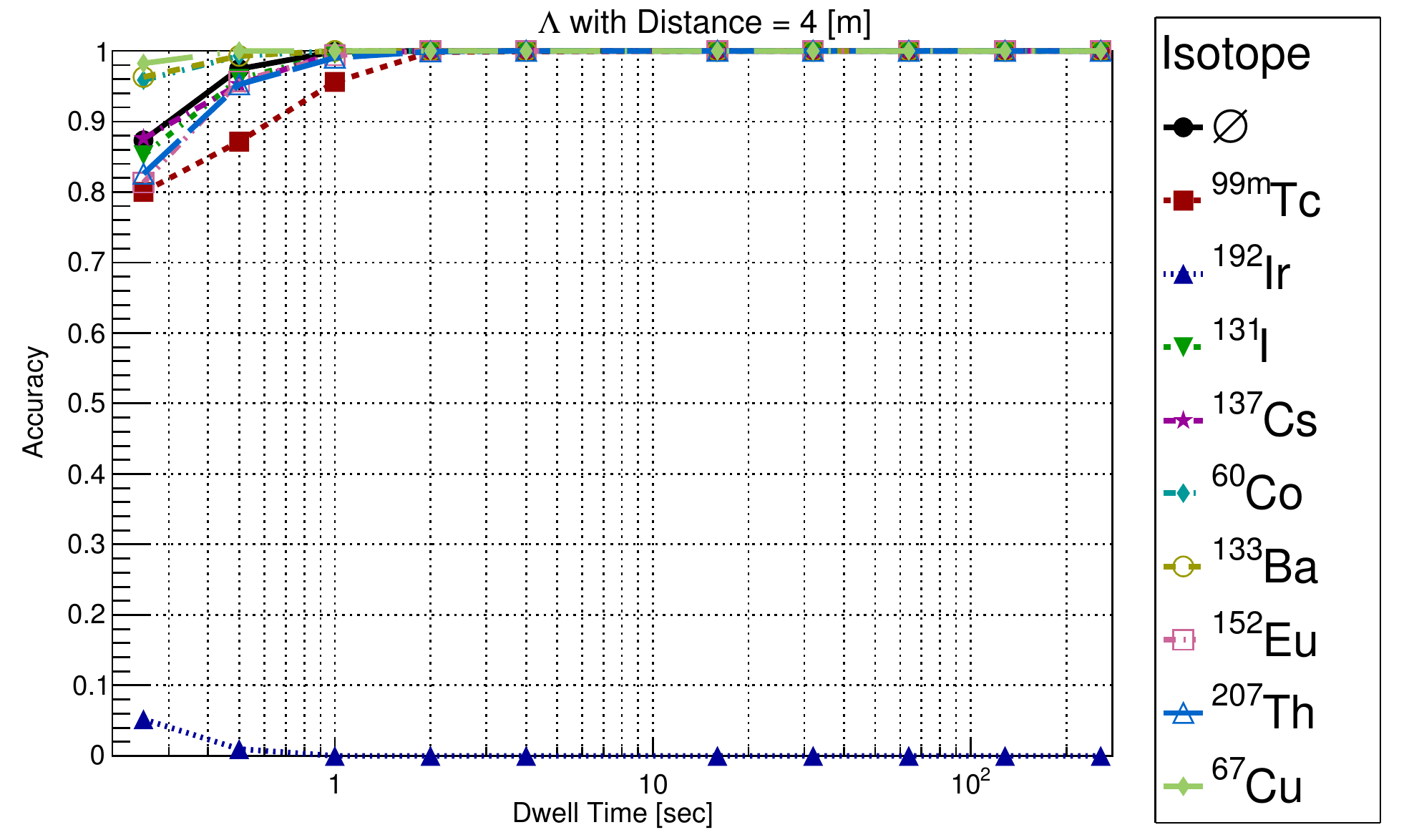} \\
      \includegraphics[width=0.48\linewidth]{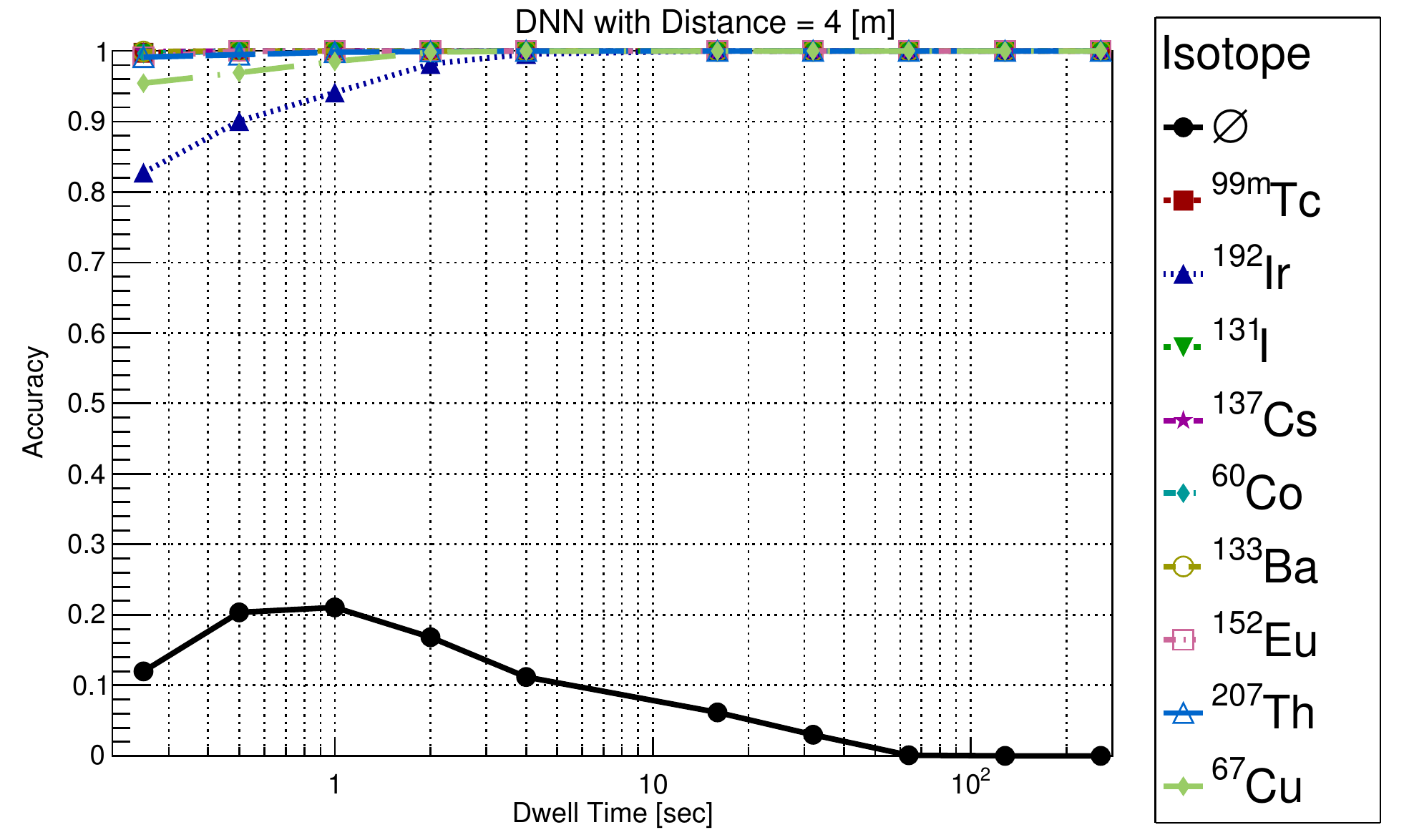} &
      \includegraphics[width=0.48\linewidth]{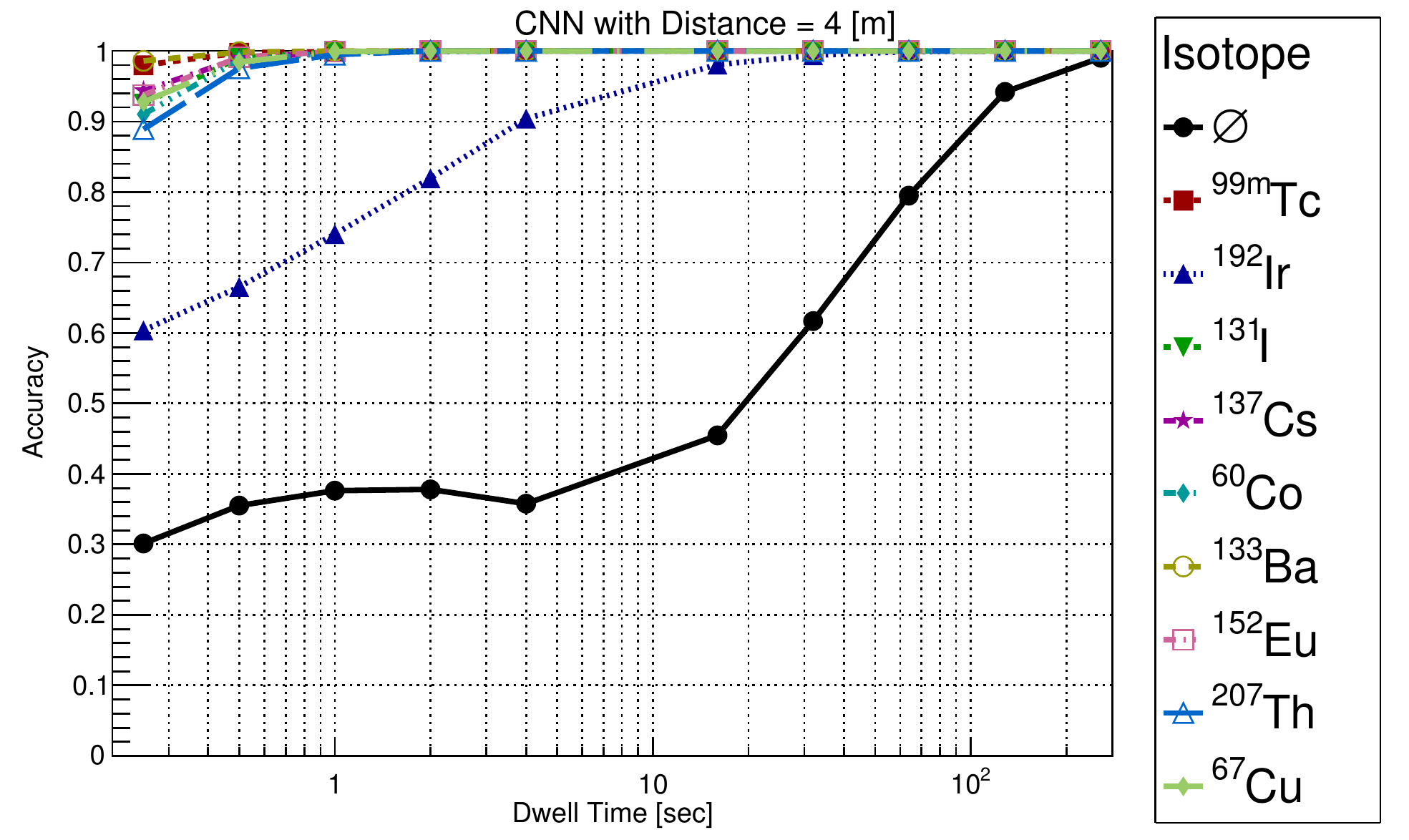} \\
    \end{tabular}    
  \end{center}
  \caption[example]{
    \label{fig:AccuracyD04-isos} 
    The class-wise accuracy of the methods as a function of dwell-time at a distance of 4 meters.
  }
\end{sidewaysfigure}

The accuracy as a function of time for a distance of four meters for all methods, and seperated for
each isotope class, is shown in Figure \ref{fig:AccuracyD04-isos}. 
We can see that even for long  dwell-times and at this relatively close distance the non-adaptive 
techniques have difficulty accurately selecting $^{192}$Ir. 
We note that the DNN at this distance fails at selecting background even at long dwell-times; this
will be investigated further. 
The CNN at this distance, however, does performs notably better than the other methods.

\begin{sidewaysfigure}[ht]
  \begin{center}
    \begin{tabular}{ccc}
      \includegraphics[width=0.48\linewidth]{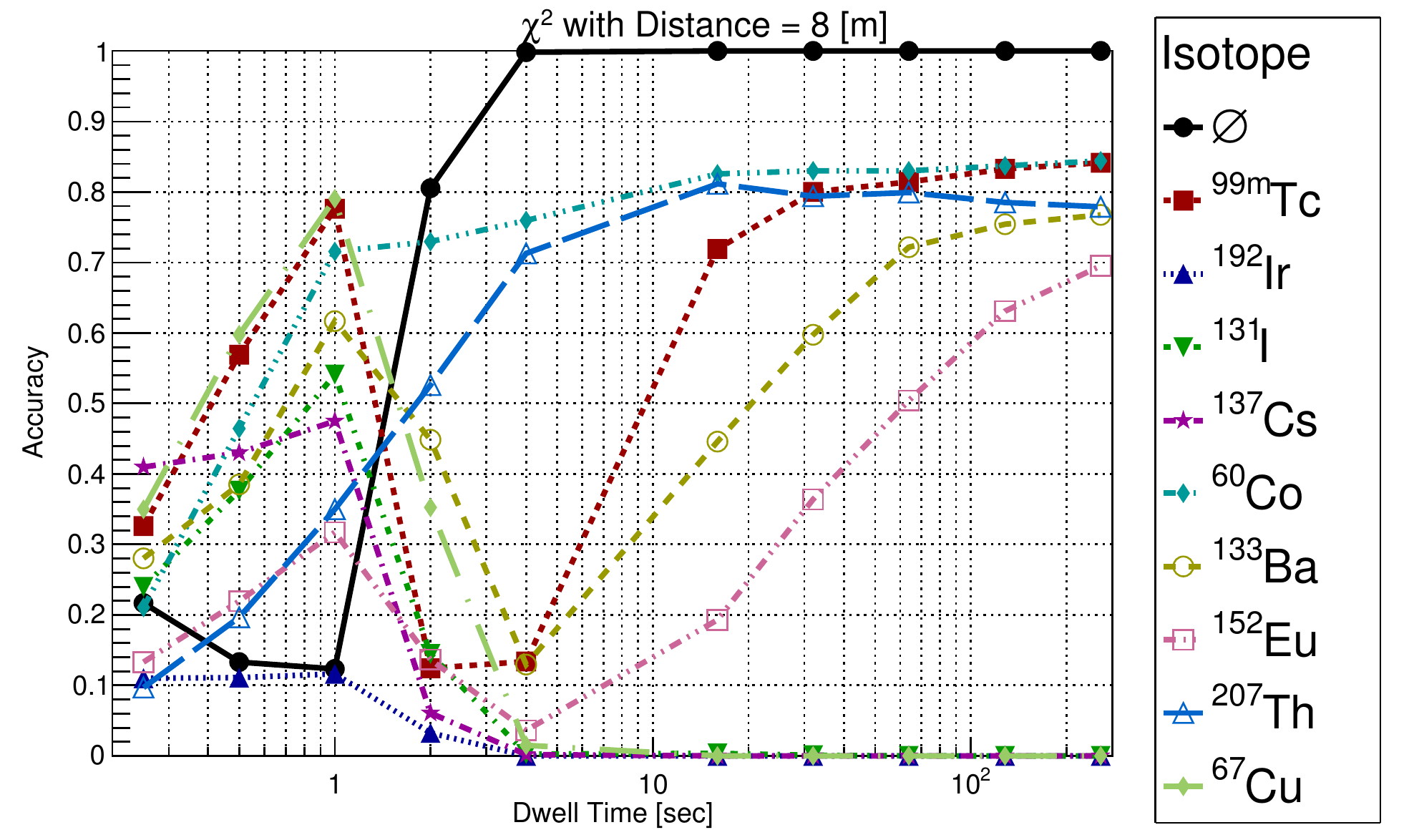} &
      \includegraphics[width=0.48\linewidth]{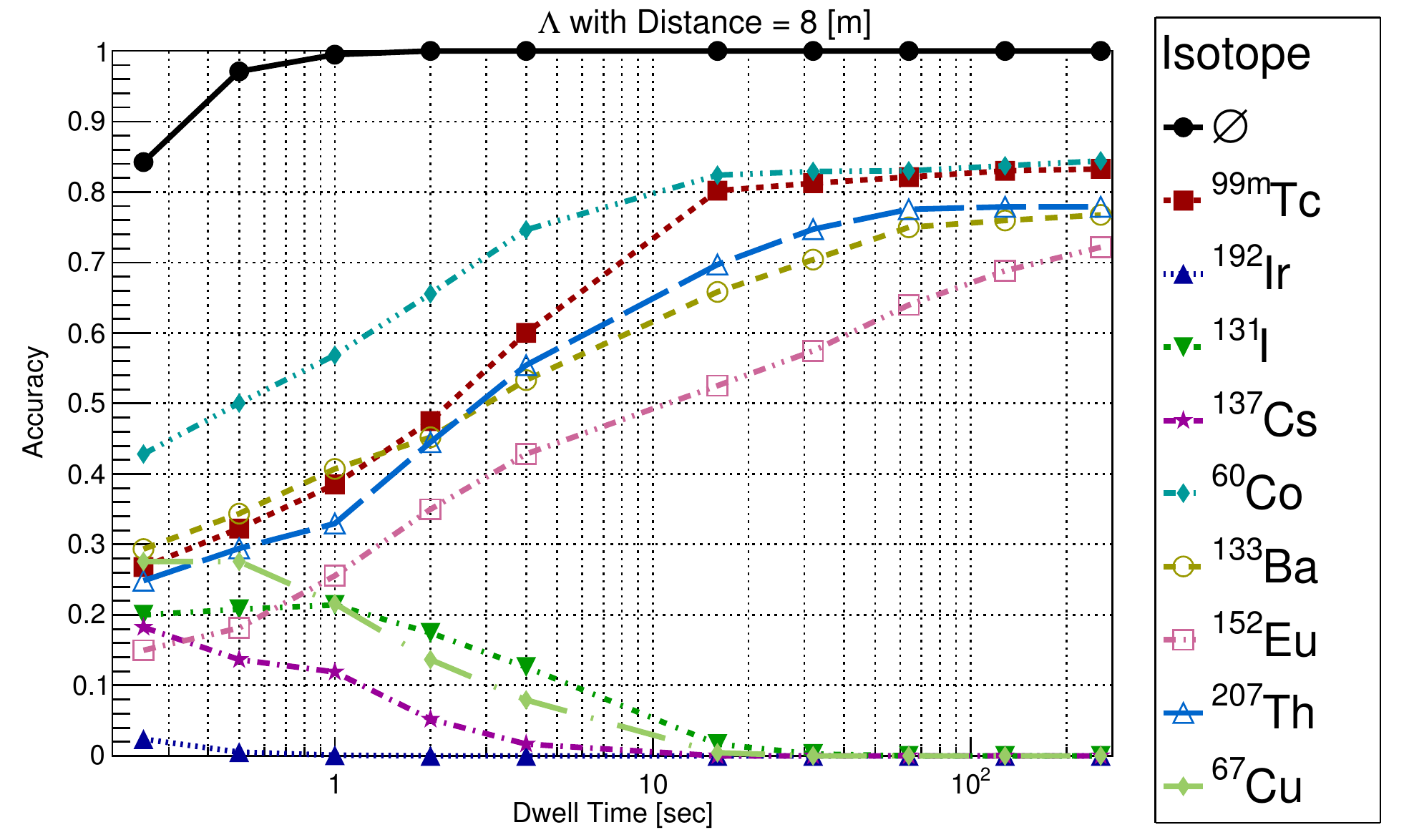} \\
      \includegraphics[width=0.48\linewidth]{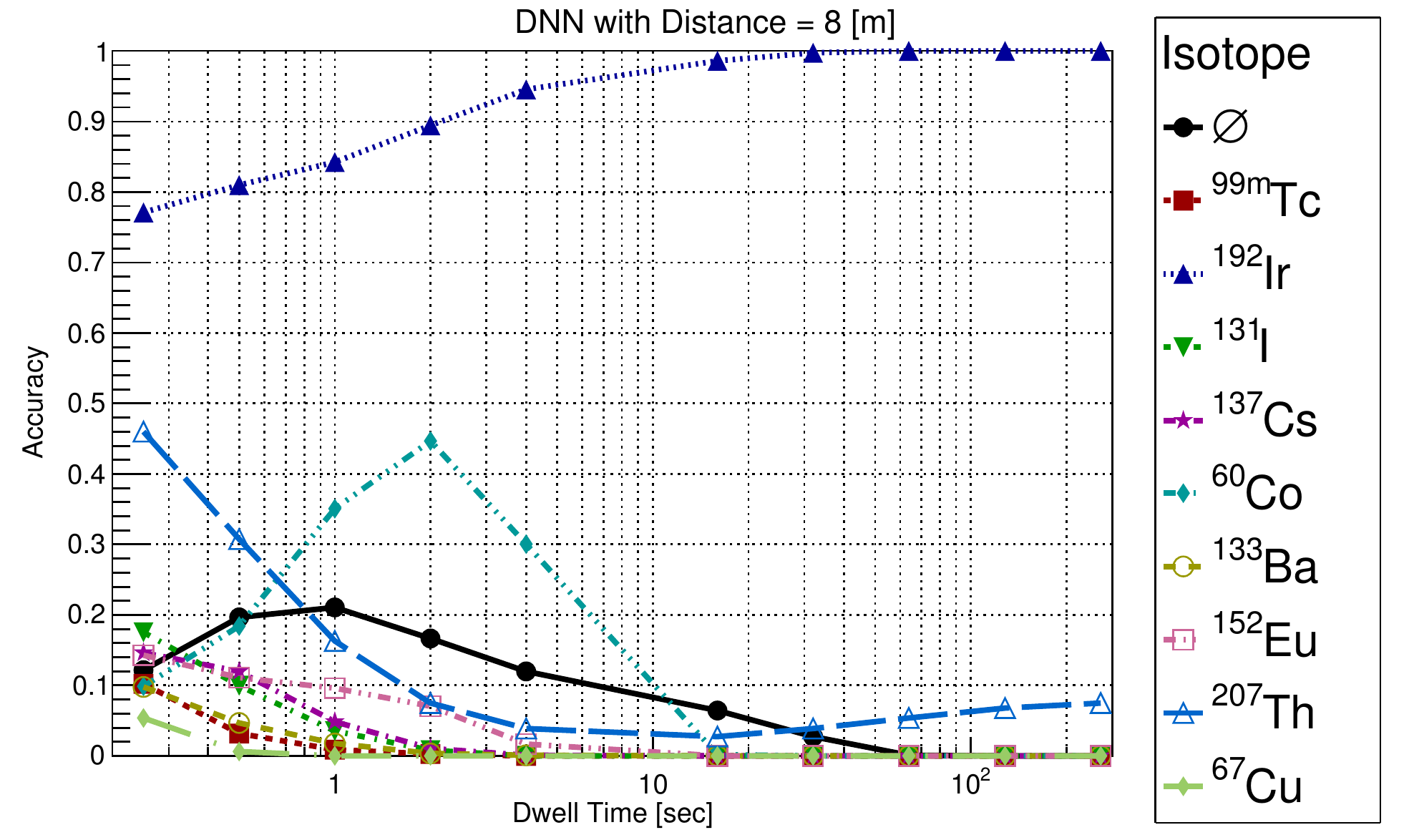} &
      \includegraphics[width=0.48\linewidth]{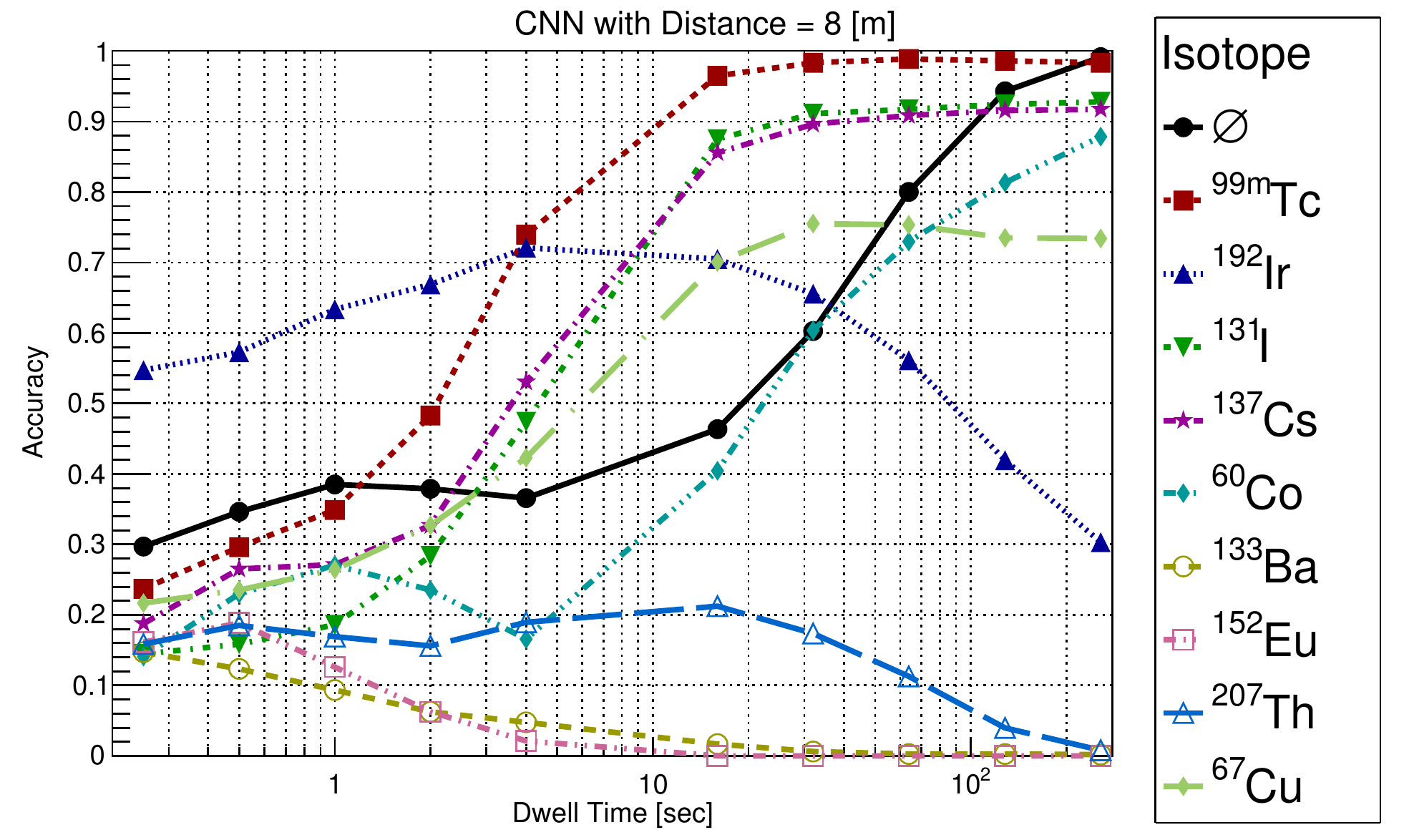} \\
    \end{tabular}    
  \end{center}
  \caption[example]{
    \label{fig:AccuracyD08-isos} 
    The class-wise accuracy of the methods as a function of dwell-time at a distance of 8 meters.
  }
\end{sidewaysfigure}

The spectra simulated at 8 meters most drastically shows the generalization performance of each
neural network.
The bottome left panel of Figure \ref{fig:AccuracyD08-isos} shows that the DNN was overfitting 
to the training dataset, misidentifying most isotopes as $^{192}$Ir. 
Because DNN's do not assume spatial structure (photopeaks, Compton continua) in data, 
this structure needs to be learned during training. 
If the training dataset is not diverse enough, the DNN may easily overfit to noise in a signal. 
This could show an inherent flaw in using an adaptive method that does not assume local structure
in the data. 
The top left and right panels of Figure \ref{fig:AccuracyD08-isos} show the non-adaptive methods
commonly misidentifying $^{192}$Ir, $^{131}$I, $^{137}$Cs, and $^{67}$Cu as background in spectra
measured at 8 meters.

Because CNN's do use the signals local spatial structure, they my be more robust when
generalizing to new signals. 
The bottom right panel of Figure \ref{fig:AccuracyD08-isos} shows that despite the differences
between the training dataset, the CNN does generalize well to the spectra measured at 8
meters. 

\begin{sidewaysfigure}[ht]
  \begin{center}
    \begin{tabular}{ccc}
      \includegraphics[width=0.48\linewidth]{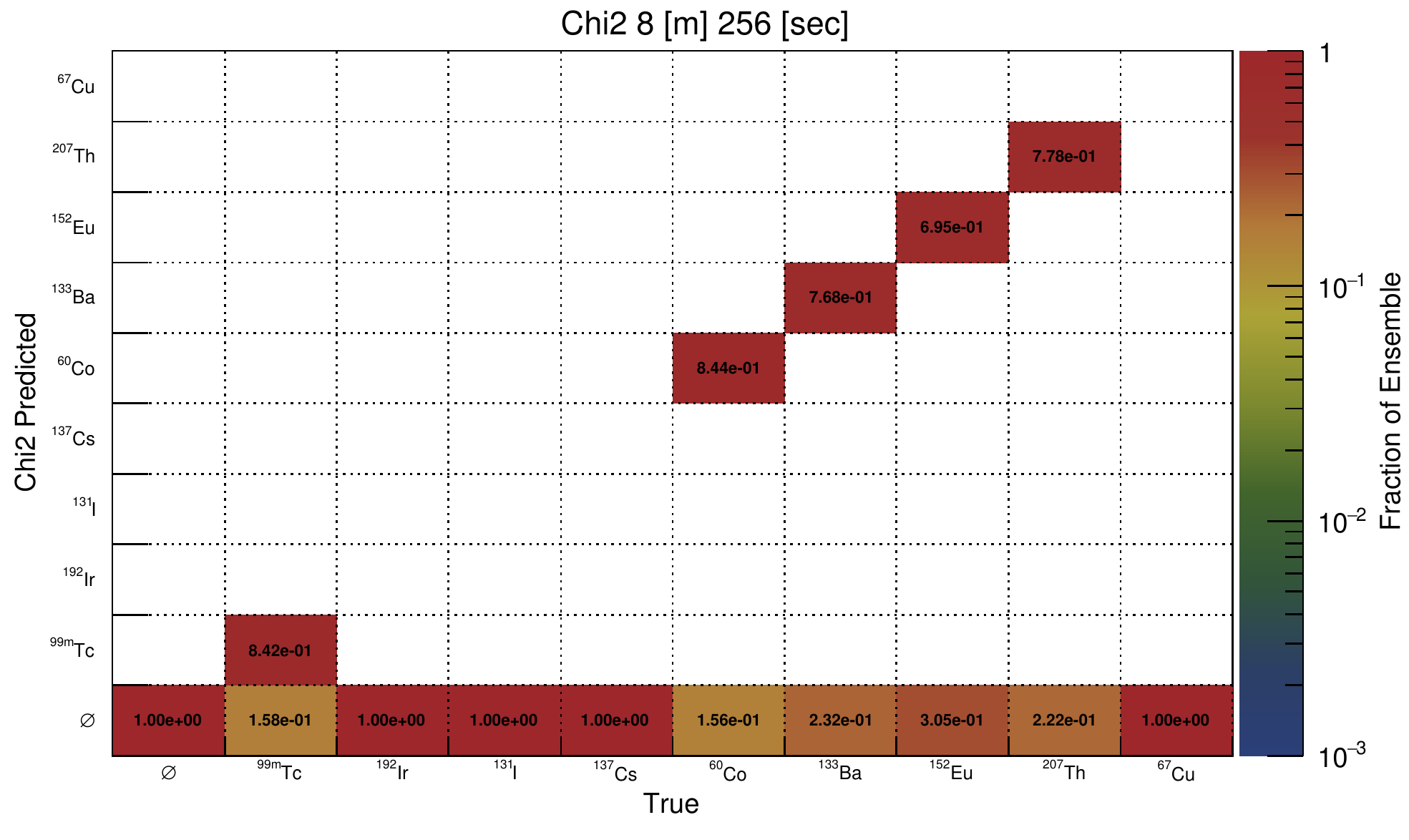} &
      \includegraphics[width=0.48\linewidth]{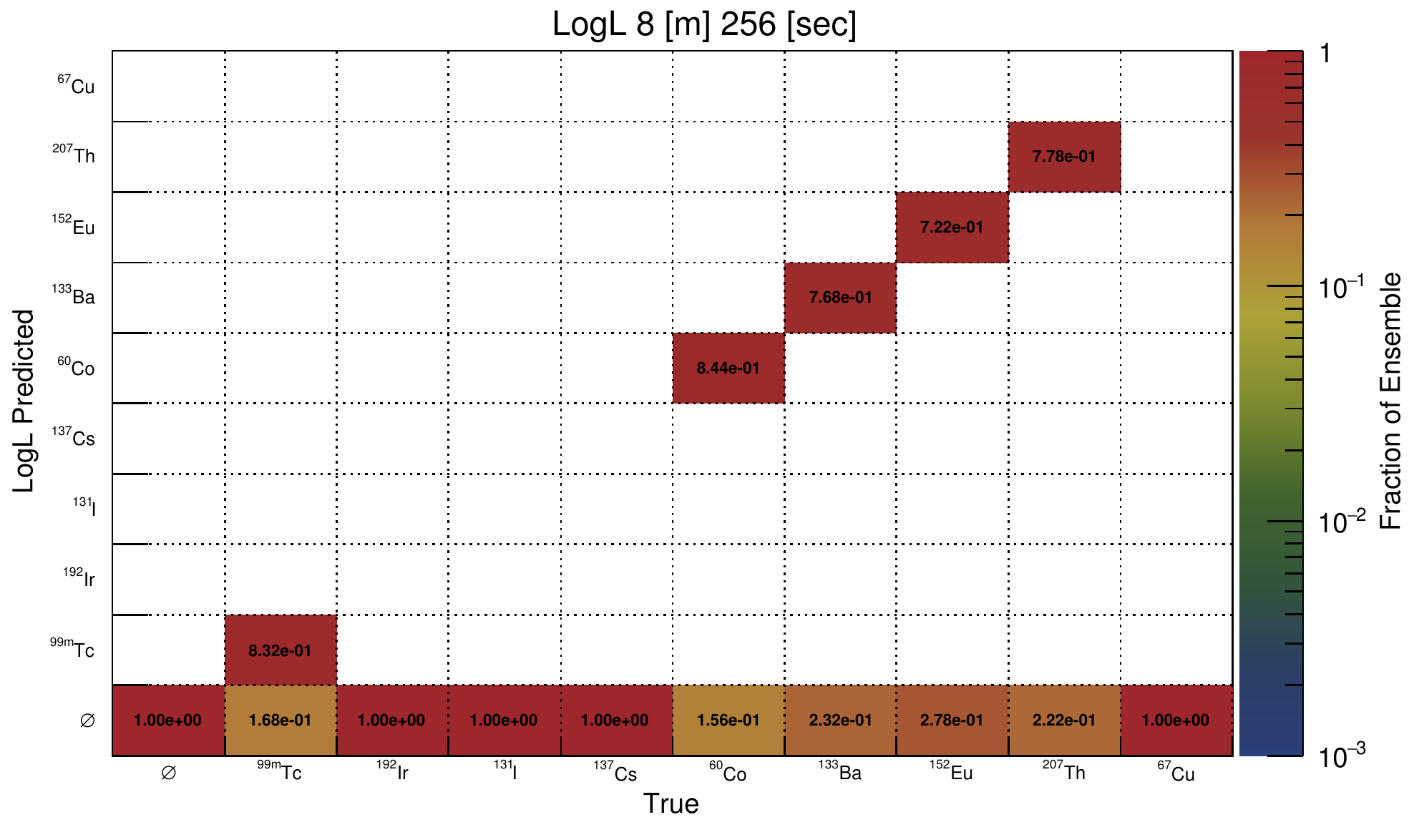} \\
      \includegraphics[width=0.48\linewidth]{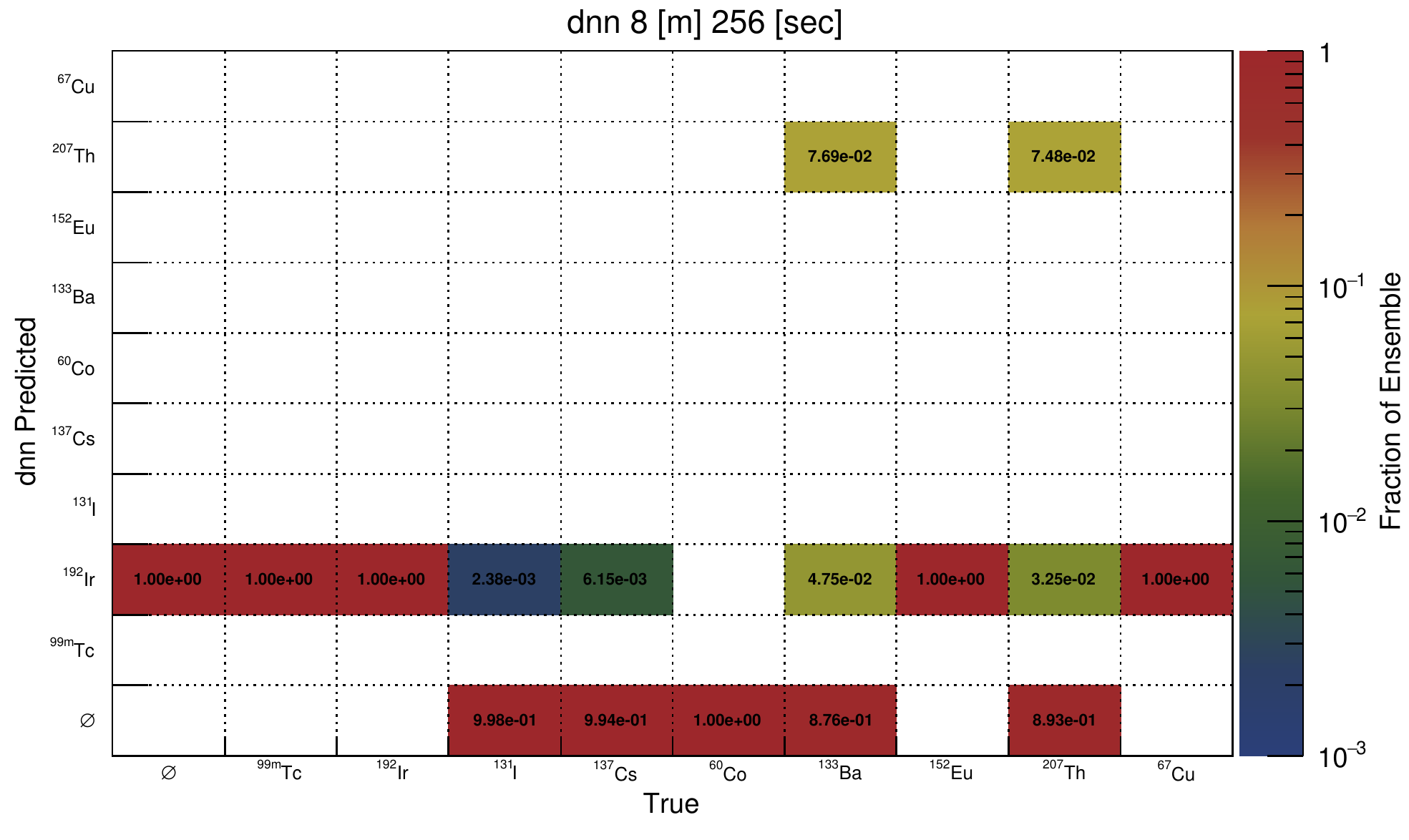} &
      \includegraphics[width=0.48\linewidth]{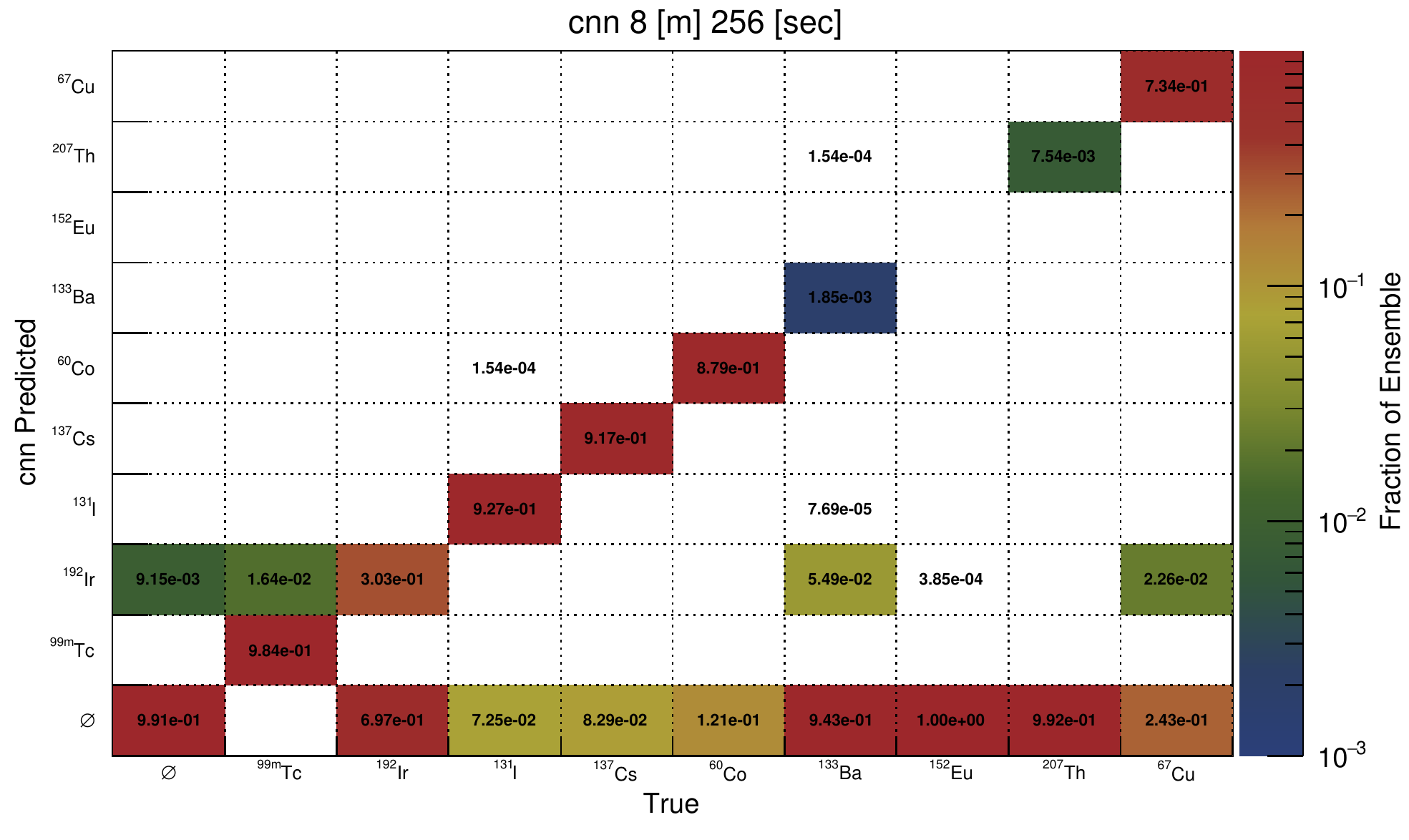} \\
    \end{tabular}    
  \end{center}
  \caption[example]{
    \label{fig:confusions} 
    The confusion matrix for each method at a distance of 8 meters and dwell-time of 256 seconds.
  }
\end{sidewaysfigure}

Finally, in Figure \ref{fig:confusions} we show the full confusion matrices for each method at a
distance of 8 meters and a dwell-time of 256 seconds. 
We note that while all methods tend to favor mis-classifying all isotopes as background, 
the DNN show the most potential for extracting accurate information when the signal strength 
is subtle, but clearly non-zero.

\clearpage

\clearpage
\pagebreak
\section*{ACKNOWLEDGMENTS}   
This work was done by Mission Support and Test Services, LLC, under
Contract No. DE-NA0003624 with the U.S. Department of Energy: DOE/NV/03624--0428.
This work was also funded by the Consortium for Verification Technology under 
Department of Energy National Nuclear Security Administration award number 
DE-NA0002534.

\bibliography{article}       
\bibliographystyle{spiebib}  

\end{document}